\documentstyle[12pt,epsf]{article}
\newcommand{\r}[1]{(\ref{#1})}
\newcommand{\ice}[1]{\relax}

\def\bbuildrel#1_#2^#3%
{\mathrel{\mathop{\kern 0pt#1}\limits_{#2}^{#3}}}

\newcommand{\as}{\alpha_s}
\newcommand{\beq}{\begin{equation}}
\newcommand{\eeq}{\end{equation}}
\begin{document}
\newcommand{\Da}{{\Delta}^*_{as}}
\newcommand{\unl}{\underline}
\newcommand{\fos}[2]{\>\>\mathop{#2}^{{}\atop{#1}}{}}
\newcommand{\aseqmy}[1]{{=\!=\!=\!=\!=}_%
{\!\!\!\!\!\!\!\!\!\!\!\!\!\!\!\!\!\!\!\!\!\!\!\!\!\!%
{{}_{\displaystyle #1}}}\>\>}
\newcommand{\Prm}{{\rm P}}
\newcommand{\Srm}{{\rm S}}
\newcommand{\Wrm}{{\rm W}}
\newcommand{\Vrm}{{\rm V}}
\newcommand{\Lrm}{{\rm L}}
\newcommand{\srm}{{\rm s}}
\newcommand{\dif}{{\rm d}}
\newcommand{\ex}{{\rm e}}
\newcommand{\MMS}{${\ovl{\mbox{ \rm mw}}}$}
\newcommand{\MS}{{{\mbox{ \rm mw}}}}
\newcommand{\wtd}{\widetilde}
\newcommand{\prd}{\partial}
\newcommand{\bea}{\begin{eqnarray}}
\newcommand{\eea}{\end{eqnarray}}
\newcommand{\non}{\nonumber}
\newcommand{\EQN}{\label}
\newcommand{\ovl}{\overline}
\newcommand{\ba}{\begin{array}}
\newcommand{\ea}{\end{array}}
\newcommand{\al}{\alpha}
\newcommand{\la}{\lambda}
\newcommand{\La}{\Lambda}
\newcommand{\be}{\beta}
\newcommand{\hA}{\hat A}
\newcommand{\hB}{\hat B}
\newcommand{\hO}{\hat O}
\newcommand{\Phibf}{{\bf \Phi}}
\newcommand{\phibf}{\underline{\bf \phi}}
\newcommand{\vphi}{\varphi}
\newcommand{\vphibf}{\underline{\bf \varphi}}
\newcommand{\Z}{{\cal Z}}
\newcommand{\Lc}{{\cal L}}
\newcommand{\I}{\mbox{I}}
\newcommand{\II}{\mbox{II}}
\newcommand{\Nbf}{{\rm{\bf  N}}}

\newcommand{\Wu}{W_{\!{}_U}}
\newcommand{\Fu}{F_{\!{}_U}}
\newcommand{\dsp}{\displaystyle}
\newcommand{\eol}{\endgraf\noindent}
\newcommand{\llp}{{}'\!\!\!\!}
\newcommand{\lp}{{}'}
\newcommand{\aseq}{\mathop{=\!\!=\!\!=}}
\newcommand{\intm}{\mathop{\int}_{\M}}
\newcommand{\inte}{\mathop{\int}_{\E}}
\newcommand{\Kep}{{\hat{K}}_\ep}
\newcommand{\Ko}{{\hat{K}}_\omega}
\newcommand{\J}{{\cal {J}}}
\newcommand{\T}{{\cal {T}}}
\newcommand{\R}{{\cal {R}}}
\newcommand{\sse}{\subseteq}
\newcommand{\lei}{\widetilde {\le}}
\newcommand{\gei}{\widetilde {\ge}}
\newcommand{\li}{\widetilde <}
\newcommand{\gi}{\widetilde >}
\newcommand{\Wt}{\widetilde W}
\newcommand{\Dtb}{\overline{\widetilde\D}}
\newcommand{\Dt}{\widetilde \D }
\newcommand{\ct}{\widetilde c}
\newcommand{\Ct}{\widetilde C}
\newcommand{\Rt}{\widetilde R}
\newcommand{\up}{\!\uparrow}
\newcommand{\g}{\gamma}
\newcommand{\G}{\Gamma}
\newcommand{\Ru}{R_{{}_U}}
\newcommand{\Rum}{R_U^{-1}}
\newcommand{\Rm}{R^{-1}}
\newcommand{\D}{\Delta}
\newcommand{\dD}{\dot{\Delta}}
\newcommand{\dDu}{\dot{\Delta}_{{}_U}}
\newcommand{\Dm}{\Delta^{-1}}
\newcommand{\Du}{\Delta_{{}_U}}
\newcommand{\Di}{\Delta_{{}_I}}
\newcommand{\Ri}{R_{{}_I}}
\newcommand{\ep}{\epsilon}
\newcommand{\de}{\delta}
\newcommand{\Mbf}{{\bf M}}
\newcommand{\x}{{\bf x}}
\newcommand{\e}{{\bf e}}
\newcommand{\f}{{\bf f}}
\newcommand{\m}{{\bf m}}
\newcommand{\p}{{\bf p}}
\newcommand{\q}{{\bf q}}
\newcommand{\z}{{\bf z}}
\newcommand{\k}{{\bf k}}
\newcommand{\LG}{{\cal{L}}_{\Gamma}}
\newcommand{\VG}{{\cal{V}}_{\Gamma}}
\newcommand{\EG}{{\cal{E}}_{\Gamma}}
\newcommand{\Lg}{{\cal{L}}_{\gamma}}
\newcommand{\Vg}{{\cal{V}}_{\gamma}}
\newcommand{\Eg}{{\cal{E}}_{\gamma}}
\newcommand{\M}{{\cal{M}}}
\newcommand{\V}{{\cal{V}}}
\newcommand{\E}{{\cal{E}}}
\newcommand{\Pc}{{\cal{P}}}
\newcommand{\C}{{\cal{C}}}
\newcommand{\F}{{\cal{F}}}
\newcommand{\N}{{\cal{N}}}
\newcommand{\Mc}{{\cal{M}}}
\newcommand{\Qbf}{{\bf{Q}}}
\newcommand{\Cbf}{{\bf{C}}}
\newcommand{\gs}{\widetilde>}
\newcommand{\ls}{\widetilde<}
\newcommand{\tld}{\sim}
\newcommand{\gm}{\gamma_m}
\newcommand{\gaam}{\gamma^{AA}_m}
\newcommand{\gaaq}{\gamma^{AA}_q}
\newcommand{\gssm}{\gamma^{SS}_m}
\newcommand{\gssq}{\gamma^{SS}_q}
\newcommand{\gppm}{\gamma^{PP}_m}
\newcommand{\gppq}{\gamma^{PP}_q}
\newcommand{\dmu}{\mu^2\frac{d}{d\mu^2}}
\newcommand{\msbar}{\overline{\mbox{MS}}}
\newcommand{\msszero}{\langle  O_2  \rangle |_{\mu_0}}
\newcommand{\muuzero}{\langle   m   \ovl u u \rangle}
\newcommand{\GGzero}{\langle   O_1  \rangle |_{\mu_0}}
\newcommand{\ASa}{a}
\newcommand{\ASzero}{a_0}

\begin{titlepage}
\noindent
%
%
\renewcommand{\thefootnote}{\fnsymbol{footnote}}
\hfill TTP94--08\footnote{The complete postscript
file of this preprint, including
figures, is available via anonymous ftp at
ttpux2.physik.uni-karlsruhe.de (129.13.102.139) as /ttp94-08/ttp94-08.ps}\\
\renewcommand{\thefootnote}{\arabic{footnote}}
\addtocounter{footnote}{-1}
\mbox{}
\hfill  June  1994   \\   %
%
%
\vspace{0.5cm}
\begin{center}
\begin{Large}
\begin{bf}
Quartic Mass Corrections to $R_{had}$
   \\
  \end{bf}
  \end{Large}
%
%
  \vspace{0.8cm}
  \begin{large}
K.G.Chetyrkin\footnote{On leave from Institute for Nuclear Research
of the Russian Academy of Sciences, Moscow, 117312, Russia},
J.H.K\"uhn\footnote{Work supported by BMFT contract 056KA93P}\\[3mm]
    Institut f\"ur Theoretische Teilchenphysik\\
    Universit\"at Karlsruhe\\
    Kaiserstr. 12,    Postfach 6980\\[2mm]
   D-76128 Karlsruhe, Germany\\
  \end{large}
%
%
  \vspace{1cm}
  {\bf Abstract}
\end{center}
\begin{quotation}
\noindent
The influence of nonvanishing quark masses on the total cross section
in electron positron collisions and on the $Z$ decay rate is calculated.
The corrections are expanded in $m^2/s$ and $\as$. Methods similar to
those applied for the quadratic mass terms allow to derive the
corrections  of order $\as m^4/s^2$ and $\as^2m^4/s^2$. Coefficients
which depend logarithmically on $m^2/s$ and which
cannot be absorbed in a running quark mass arise in order $\as^2$.
The implications of these
results on electron positron annihilation cross sections
at LEP and at lower energies in particular between the
charm and the bottom threshold and for energies several GeV above the
$b\bar b$ threshold are discussed.

\end{quotation}
\end{titlepage}

\section{Introduction}
The precise measurement of the cross section for hadron production
in electron positron annihilation leads to an
accurate determination of the strong coupling constant $\as$. The method
is free from a variety of theoretical uncertainties and ambiguities
which beset many other observables of (still) superior statistical
accuracy. The extraction of $\as$ from the ratio $R$ between the
rates for hadron and lepton pair production is based on the
calculation of perturbative QCD corrections. These are available
up to third order in $\as$ in the massless limit \cite{Chet,Gor}.
Mass corrections were calculated for small $m^2/s$ up to
order  $\as^3$ for vector and up to order $\as^2$ for axial
vector current induced reactions
\cite{ChetKuhn90,ChetKuhnKwiat92}.
In these papers it was furthermore demonstrated
that all large logarithms which arise in the expansion coefficients can
be absorbed in the running quark mass. The same  comment applies  to the
contribution of ``singlet" terms which are relevant for the
axial part of the $Z$ decay rate
\cite{KniehlKuhn90,CK4,ChetTar93,ChetKwiat93}.

In principle one would of course prefer to control the mass dependence
of the cross section for arbitrary $m^2/s$. To date this has been
achieved in first order in $\as$ only. Higher orders in $\as$ were
calculated for the first term in the
$m^2/s$ expansion, allowing for calculational
techniques based on massless propagator type integrals. In this work
this program will be pursued further and $m^4/s^2$ terms will be
calculated up to order $\as^2$.  These terms are fairly unimportant
for $Z$ decays, however, they will become relevant
in  the low energy region. Examples are measurements
at a $B$ meson factory just
below or several GeV above the $b\bar b$ threshold,
or alternatively around 5~GeV, an energy
region that could be explored at the BEPC storage ring.

The $\as^2$ calculation presented below is based on
\cite{ChetGorSpi85,CheSpi88}.
There
the operator product expansion of two point correlators was studied up
to two loops, including therefore terms of first order in $\as$. The
expansion included power law suppressed terms up to operators of
dimension four, which are induced by nonvanishing quark masses.
Renormalisation group arguments, similar to those employed already in
\cite{ChetKuhn90,ChetKuhnKwiat92},
will allow to deduce the $\as^2 m^4$ terms. The
calculation is performed for vector and axial vector
current correlators. The first one is of course
relevant for electron positron annihilation into heavy quarks at
arbitrary energies, the second one for $Z$ decays into $b$ quarks and
for top production at a future linear collider.
In order  $\as$ the $m^4$ terms can also be obtained from the Taylor
expansion of the complete answer \cite{schw,Rein}. This
provides not only a (fairly trivial) cross check of the $\as$ term. The
comparison with the complete answer in first order $\as$
may also indicate to which extent
$m^2$ plus $m^4$ terms lead to a reliable estimate of the complete
answer in order $\as^2$.

\section{Mass terms  of first order in $\as$}
QCD corrections to vector and axial current correlators
in order $\as$ and for arbitrary $m^2/s$ were  derived in
\cite{schw}. Compact formulae are given in \cite{Rein}. These
are conventionally expressed in terms of the pole mass
denoted by $m$ in the following. It is straightforward to expand these
results in $m^2/s$ and one obtains -- in obvious notation  --
\begin{eqnarray}
R_V&=&1 - 6\frac{m^4}{s^2} - 8\frac{m^6}{s^3}\\
&& +\frac{\as}{\pi}\biggl[1+12 \frac{m^2}{s}
+ \left( 10 - 24\log (\frac{m^2}{s})\right)\frac{m^4}{s^2}
                    \nonumber\\
   &&  -\frac{16}{27}\left(47 + 87 \log (\frac{m^2}{s})\right)
   \frac{m^6}{s^3}\biggr], \nonumber\\
R_A&=&1 - 6\frac{m^2}{s} + 6\frac{m^4}{s^2} + 4\frac{m^6}{s^3}\\
&& +\frac{\as}{\pi}\biggl[
     1-\left(6+12\log( \frac{m^2}{s})\right)\frac{m^2}{s}+
       \left(-22 + 24\log (\frac{m^2}{s})\right)\frac{m^4}{s^2}
                    \nonumber\\
   &&  +\frac{8}{27}\left(41 + 42 \log (\frac{m^2}{s})\right)
   \frac{m^6}{s^3}\biggr]\nonumber
. 
\label{1}
\end{eqnarray}

\begin{figure}
\begin{center}
\epsfxsize=12.0cm
\leavevmode
\epsffile[130 300 460 525]{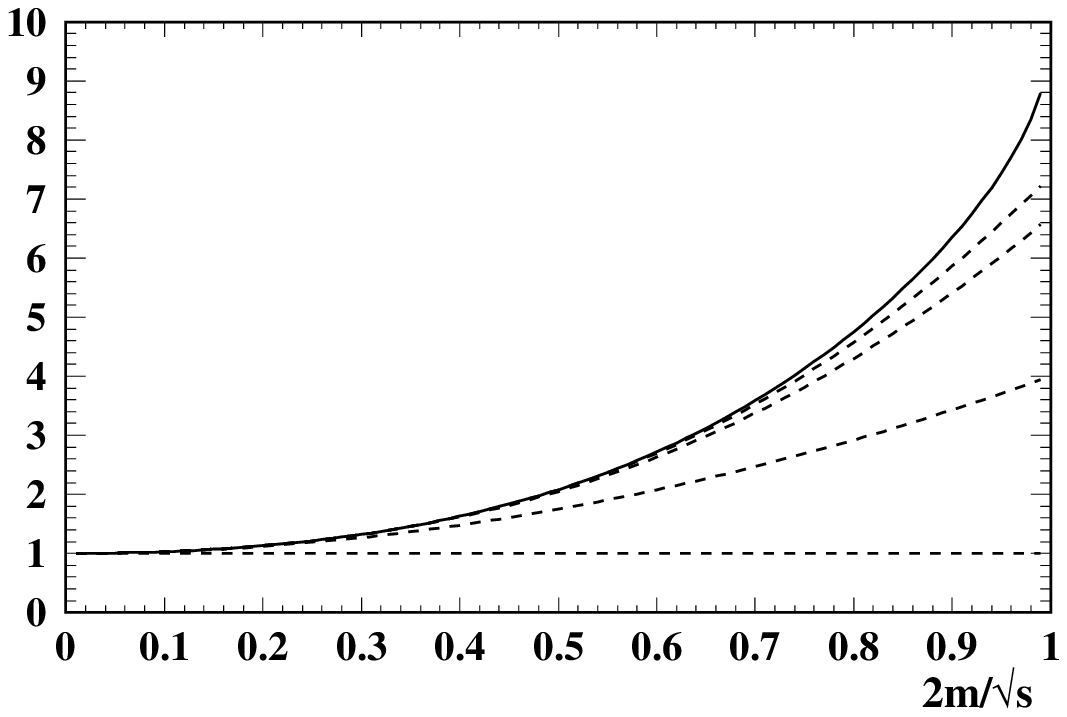}\\
\epsfxsize=12.0cm
\leavevmode
\epsffile[130 300 460 525]{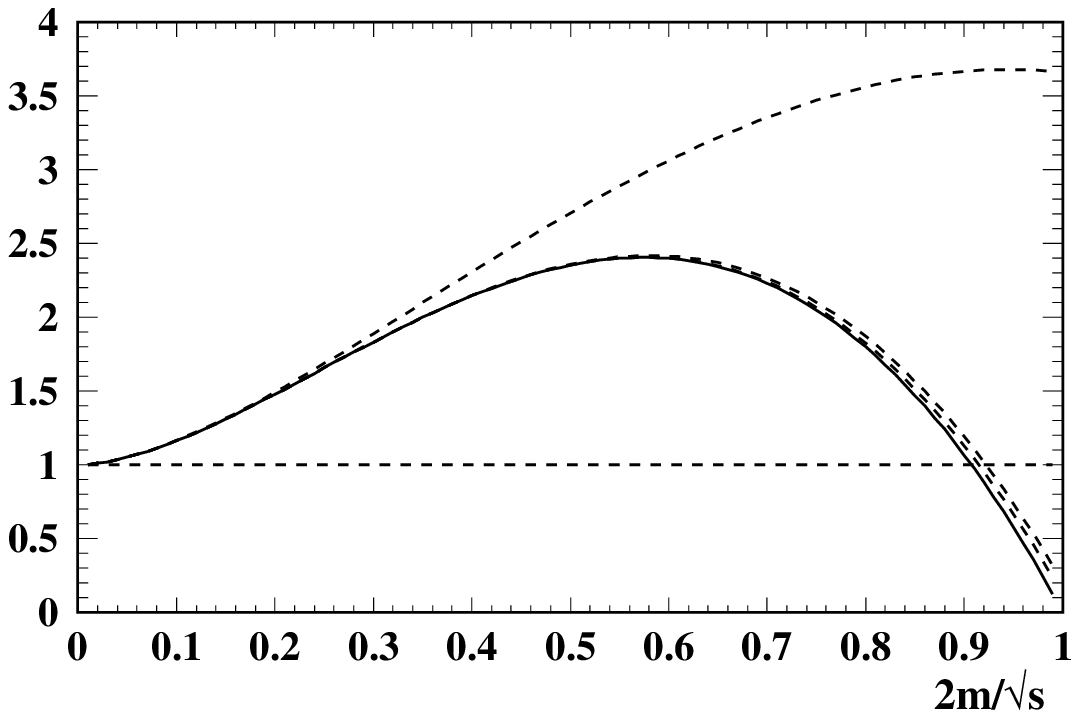}
\caption{\label{kvexp}Comparison between the
complete ${\cal O}(\alpha_s)$ correction
function (solid line) and approximations of increasing order
(dashed lines) in $m^2$ for vector (upper Figure)
and axial vector current (lower Figure) induced rates.}
\end{center}
\end{figure}

The approximations to
the correction function for the vector current correlator
(including successively higher orders and without the
factor $\alpha_s/\pi$) are compared to the full result in
Fig.\ref{kvexp}.
For high energies, say for $2 m_b/\sqrt{s}$ below 0.3, an
excellent approximation is provided by the constant plus the $m^2$ term.
In the region of $2m/\sqrt{s}$ above 0.3 the $m^4$ term becomes
increasingly important. The inclusion of this term improves the
agreement significantly and leads to an excellent approximation even up
to $2m/\sqrt{s}\approx 0.7$ or 0.8. For the narrow region between 0.6
and 0.8 the agreement is further improved through the $m^6$ term.
Higher powers in $m$ which can be calculated in a straightforward way,
lead  only to a modest further improvement in the region close to
threshold, say above 0.8. This region with its behaviour governed by the
Coulomb singularity requires therefore a different treatment. The same
considerations apply for the axial current correlator
(Fig.\ref{kvexp}).

Inclusion of $m^2$ and $m^4$ terms hence allows to extend the QCD
prediction
from the high energy limit down to fairly low center of mass
energy values, perhaps 4 to 5 GeV above the $b\bar b$ threshold, or from
about 2 GeV above the charm threshold up to just below $\Upsilon(4S)$.
Similarly to the  discussion in \cite{ChetKuhn90,ChetKuhnKwiat92}
for the $m^2$ terms,
the logarithms accompanying the $m^4$ terms
can also be absorbed through a redefinition of $m$ in terms of
the $\overline{MS}$ mass \cite{Tarrach81,Narison87}
at scale $s$
\beq
m^2=\ovl{m}^2(s) (1+ \frac{\as}{\pi}(-2\log m^2/s +8/3))
\eeq
and one obtains\footnote{The $m^4$ terms given below are in disagreement
with those quoted in \cite{Narison}.}($\ovl{m} \equiv \ovl{m}(s)$)
\begin{eqnarray}
R_V& =& 1 - 6\frac{\ovl{m}^4}{s^2} -8\frac{\ovl{m}^6}{s^3}\\
&&+\frac{\as}{\pi}\left
    [1 + 12\frac{\ovl{m}^2}{s} - 22 \frac{\ovl{m}^4}{s^2}
-  \frac{16}{27}\left(6\log( \frac{\ovl{m}^2}{s})  + 155\right)
  \frac{\ovl{m}^6}{s^3}
   \right], \nonumber\\
R_A& =& 1 - 6 \frac{\ovl{m}^2}{s} + 6\frac{\ovl{m}^4}{s^2}
+  4\frac{\ovl{m}^6}{s^3}\\
&& +    \frac{\as}{\pi}\left
   [1 - 22\frac{\ovl{m}^2}{s} + 10 \frac{\ovl{m}^4}{s^2}
 + \frac{8}{27}\left(- 39 \log( \frac{\ovl{m}^2}{s})
+  149\right)\frac{\ovl{m}^6}{s^3}
\right]
. \nonumber
\end{eqnarray}
This resummation is possible for the second and fourth powers of $m$
in first
order $\as$ and in fact for $m^2$ corrections to $R_V$ and $R_A$ in all
orders of $\as$. However, logarithmic terms persist in the
$m^4$ corrections, starting from ${\cal O}(\as^2)$,
as discussed in the next section. Higher order terms in the mass
expansion, starting from $m^6$, evidently exhibit superficially
``incurable'' logs already in order $\as$. In fact, these logs may be
also summed up (see \cite{BroadGen84,CheSpi88}
and also \cite{CK94}).
Unlike the quadratic
mass term, the resulting expression will contain some explicit
dependence on the strong coupling constant $\as(m_q)$ taken at the
scale equal to the quark mass.  The numerical effect of such a
summation proves to be rather small unless the ratio $s/m_q^2$ is
chosen   too large.

\section{Corrections of order $\as^2m^4$}

Motivated by the fact that the first few terms in the $m^2/s$ expansion
provide already an excellent approximation to the complete answer in
order $\as$, we now proceed to the evaluation of the three loop
corrections. The calculation makes use of the properties of the operator
product expansion as discussed in \cite{ChetGorSpi85,CheSpi88}.
To allow for a comprehensive presentation, the relevant formulae are
repeated whenever necessary.
\vspace{3mm}

\noindent
{\sl 1. Operator Product Expansion (OPE)}
\vspace{3mm}

\noindent
The calculation is based on the operator product expansion of
the T-product of two vector or axial vector currents
\beq
T_{\mu\nu}(q,J)= i\int T(J_\mu(x) J^+_\nu(0))\ex^{iqx} \dif x,
\EQN{141}
\eeq
with
$J_\mu=\ovl u\g_\mu d $ (or  $J_\mu=\ovl u\g_\mu\g_5 d $).
Here $u$ and $d$ are just two generically different quarks with
masses $m_u$ and $m_d$. Quarks which are not coupled to the external
current will influence the result in order $\as^2$ through their
coupling to the gluon field.
\ice{Only one such quark is considered and its
mass will be denoted generically by $m_s$.
}
The result may be immediately transformed to the case of the
electromagnetic current of a heavy, say, $t$ (or $b$ ) quark.

The asymptotic behaviour of this
(operator valued) function for  $Q^2 = -q^2\to\infty$  is given by
an OPE  of the form (Different powers of $Q^2$ may be studied
separately. Only terms proportional to Lorentz scalar
operators of dimension four are kept to derive the $m_q^4$ corrections as
eventually we are interested in  the function \r{141} sandwiched between
the vacuum states.)
\beq
\ba{l}\dsp
T_{\mu\nu}\bbuildrel{=}_{q^2\to\infty}^{} \frac{1}{Q^4}
\sum_n
\left\{
       (q_\mu q_\nu -g_{\mu\nu} q^2)\fos{T}{C_n}(Q^2,\mu^2,\al_s)

+q_\mu q_\nu \fos{L}{C_n}(Q^2,\mu^2,\al_s)\right\} O_n+\dots
\ea
\EQN{142}
\eeq

As demonstrated in \cite{Spi84,ChetGorSpi85}
the following operators of dimension four may in general appear at the
rhs. of \r{142},
\beq
\ba{l}\dsp
O_1=G_{\mu\nu}^2,\ O_2^{ij}=m_i\ovl q_j q_j,\  O_3^i=
\ovl q_i(i\hat{\fos{\leftrightarrow}{D}}/2-m_i) q_i,
\\
\dsp
O_4=A^a_\nu\left( \nabla^{ab}_\mu G^b_{\mu\nu}+
g\sum_i\ovl q_i\frac{\lambda^a}{2}\g_\nu q_i\right)
-\prd_\mu \ovl c^a\prd_\mu c^a,
\\ \dsp
O_5=(\nabla^{ab}_\mu \prd_\mu \ovl c^b) c^a,
\\ \dsp
O_{6}^{ij}=m_i^{2} m_j^{2},
\\ \dsp
O_6^i=m_u m_d m_i^2.
\ea
\EQN{143}
\eeq

The (gauge non-invariant) operators
$O_4$ and $O_5$ are non-physical; they appear since
the gauge invariance of the
Lagrangian is broken through gauge-fixing.
Matrix elements of these operators vanish for
physical and thus by definition
gauge invariant states. Hence they do not contribute in the calculation
below. They must, however, be taken into account in
the calculation of the coefficient functions (CF's) multiplying the
physical operators. Once these CF's
are known, one may freely ignore the  non-physical
operators.

In \cite{ChetGorSpi85}
the following results were obtained for the (transversal)
coefficient functions:
\beq
\fos{T}{C}_1=
\frac{\al_s}{12\pi}(1 + \frac{7}{6}\frac{\al_s}{\pi} ),
\EQN{144}
\eeq
\beq
\ba{l} \dsp
\sum \fos{T}{C_2^k} {O}^k_2=
\\ \dsp
-\frac{\al_s}{\pi}
\left(1+   \frac{\al_s}{\pi}
\left[
-\frac{1}{6} f L
-\frac{1}{6} f
+\frac{11}{4} L
+\frac{29}{6}
\right]
\right)
 (m_u \ovl u u+m_d \ovl d d)        \\ \dsp
\pm
\left
      (1+\frac{4\al_s}{3\pi}  \left[
                                    1+\frac{3\al_s}{4\pi}
                                         \left(
                                         -\frac{2}{9} f L
                                         -\frac{7}{27} f
                                         +\frac{11}{3} L
                                         +\frac{191}{18}
                                         \right)
                                \right]
\right)
 (m_u \ovl d d+m_d \ovl u u)
\\ \dsp
+(\frac{\al_s}{\pi})^2 \left(\frac{4}{3} \zeta(3)+\frac{L}{3}-1\right)
\sum_{i} m_i\ovl q_i q_i,
\ea
\EQN{145}
\eeq
\beq
\ba{l}         \dsp
\sum \fos{T}{C_6} O_6= \\   \dsp
\frac{3}{16 \pi^2}
\left\{
        \frac{\al_s}{\pi}
    \left(\frac{152}{9} - \frac{32}{3}\zeta(3)\right) m_u^2 m_d^2
     - \left( 2+ \frac{\al_s}{\pi}   (4 + 4L)\right) (m_u^4+m_d^4)
  \right.
  \\    \dsp
  \left.
        \mp \left[4 L+\frac{\al_s}{\pi}
       \left(
\frac{56}{3}
-16 \zeta(3)
+\frac{32}{3} L
+8 L^2
       \right)
         \right]
        (m_u^3 m_d+m_d^3 m_u)
\right\},
\ea
\EQN{146}
\eeq
where $L=\ln\mu^2/Q^2$,  $\zeta(3)=1.202\dots$, $f$ stands for
the number of flavours;
the upper  (lower) sign refers to  $\ovl u\g_\mu d$
($\ovl u\g_\mu\g_5 d$).

From these results one may find for example the CF's
for $T_{\mu\nu}(q,J)$ with
$J_\mu=\ovl b \g_\mu b$ or $\ovl b \g_\mu\g_5 b $
through the substitutions
$ d\to b, \  \  u  \to  b$
and  $m_d\to m_b, \ \  m_u \to  m_b$.

The starting point of the calculation was the nondiagonal $\bar u d$
current. Singlet contributions are therefore absent by construction.
These would contribute in order $\alpha_s^2$ to the axial part of
$\Gamma(Z\to\bar b b)$ and will not be considered in this work.
Their relative importance will be discussed in section 4.

It should be stressed that the generic OPE \r{142}
determines the asymptotic behaviour of any
Green function with an insertion of the T-product $T_{\mu\nu}(q,J)$,
in particular of the correlator
\beq
\Pi_{\mu\nu}(q) = \langle T_{\mu\nu}(q,J) \rangle
. 
\EQN{def.Pi}
\eeq
In particular,  as demonstrated in \cite{CheSpi88} the expansion
\r{142} bears all the information about the large $Q$ behaviour of
the correlator \r{def.Pi} in perturbation theory, provided the Vacuum
Expectation Values (VEV's)  of local  operators are
evaluated and renormalized
according to the  standard minimal subtraction prescription.
(No normal ordering of the operator
products!)
The above results allow to find the absorptive part of the
polarization operator  $\Pi(Q^2)$  defined in the standard manner
$$
\Pi_{\mu\nu}(q)  = -g_{\mu\nu} q^2 \Pi(Q^2) +\dots.
$$
once the
$\overline{MS}$ renormalized values for the vacuum expectations
values $\langle G^2 \rangle$ and $\langle \ovl q q \rangle$ are
known.

However,  this nice feature of the minimal subtraction scheme has its
price: In schemes without normal ordering
the renormalization properties of composite operators are
more involved. For instance,
the operator $m_i\ovl \psi_j \psi_j$  mixes
under renormalization  group transformations
with ``operators''  proportional to the unit ``operator''  times
polynomials of fourth order in the quark masses.
For our aims it will be necessary to know the anomalous dimension
matrix of all physical operators in the list \r{143}. It has been
shown in  \cite{Spi84,CheSpi88}
(for earlier related works see
\cite{Kluberg75,Nielsen75,Nielsen77,Tarrach82}) that
this matrix can be expressed, to all orders in perturbation theory,
in  terms of the QCD $\beta$-function, the quark mass anomalous
dimension $\g_m (\alpha_s)$ and the so-called  vacuum anomalous
dimension $\g_0$.  The explicit
expression for the anomalous dimension matrix  will be presented below.

In the calculation of $R(s)$ only the logarithmic parts
of the CF's need to be taken into account (the constant terms
have no absorptive parts at $s > 0$ !).
Three remarkable observations can be made:
\begin{itemize}
\item
   In first order of $\as$ only trivial operators (proportional
   to the unit operator times some combination of quark masses)
   contribute to $R(s)$ since only these depend on $L$.
   Hence $R(s)$ may be found directly from the coefficient function
   \r{146}. This fact directly leads to the absence of the mass
   logarithms in $R(s)$ in order $\as$ as the latter may appear only
   from  VEV's of non-trivial operators.
   This is in agreement with the expansion of the exact result
   discussed in section 2.
   Furthermore, the argument does not apply to terms of order $\alpha_s
    m^6$, again in agreement with the expansion calculated in section 2.

\item
    To find the correction of order
    $\alpha_s^2 m_q^4/s^2$ to  $R(s)$  there is  no need to compute
    three-loop contributions of order $\alpha_s^2$
    to the VEV's $\langle G^2\rangle$ and $\langle\ovl q q \rangle$.
    Only  already known two-loop terms of
    order $\alpha_s$ (see below) are required.

\item
  The calculations of \cite{ChetGorSpi85} were performed
  at the two-loop level and thus  their results
  as expressed by  \r{144}-\r{146} do not include
  terms of order $\as^2$ for the CF's
  $\dsp {\Cbf}^n_6$. Fortunately,
  the terms proportional
  to $L$ may be inferred from the renormalization group
  invariance in analogy to \cite{ChetKuhn90}.
\end{itemize}
\vspace{3mm}

\noindent
{\it 2. Renormalization Group (RG) Functions}
\vspace{3mm}

\noindent
The following notation will be employed:

\begin{itemize}
\item The effective {\em couplant} is defined as
$a(\mu) \equiv \frac{\dsp\as(\mu)}{\dsp\pi}$
and the number of  flavours is denoted by $f$.
\item
The $\beta$-function and the anomalous dimension of the mass
$\gm$ are defined through
\beq
\dmu \left({a(\mu)} \right) = a\beta(a) \equiv
-a\sum_{i\geq0}\beta_i\left(a\right)^{i+1},
\EQN{anom.mass.def}
\eeq
\beq
\dmu {m}(\mu) =  {m}(\mu)\gm(\as) \equiv
-{m}\sum_{i\geq0}\gm^i\left(a\right)^{i+1}.
\EQN{beta.def}
\eeq
with the expansion coefficients up to order ${\cal O}(\as^2)$
\beq
\ba{c}\dsp
\beta_0=\left(11-\frac{2}{3}f\right)/4,  \  \
\beta_1=\left(102-\frac{38}{3}f\right)/16, \\ \dsp
\beta_2=\left(\frac{2857}{2}-\frac{5033}{18}f+
\frac{325}{54}f^2\right)/64,
\ea
\EQN{beta3}
\eeq
\beq
\ba{c}\dsp
\g^0_m=1, \  \ \ \g^1_m=\left(\frac{202}{3}-\frac{20}{9}f\right)/16,
\\ \dsp \g^2_m=\left(1249 -
\left[\frac{2216}{27}+\frac{160}{3}\zeta(3)\right]
f-\frac{140}{81}f^2\right)/64.
\ea
\EQN{anom.mass3}
\eeq
\item
The anomalous dimension of the vacuum  energy
is\footnote{The leading term and the first correction in the
expression for $\g_0^{d}$ were
calculated in \cite{CheSpi88}. Recently one of us
\cite{Chet94} extended the calculation  to include
the correction of order $\as^2$.}
\beq
\g_0(a,m) = \g_0^{d}(a) \sum_i m_i^4
+
\g_0^{nd}(a) \sum_{ i\not= j} m_i^2 m_j^2
,
\eeq
with
\beq
\ba{c}
\dsp
\g_0^{d}(a) =
-\frac{3}{16\pi^2}
\left[  1 + \frac{4}{3} a
 + \left(\frac{313}{72}- \frac{5}{12} f - \frac{2}{3}\zeta(3) \right)a^2
\right]
,
\\
\dsp
\g_0^{nd}(a) = \frac{3}{8\pi^2}a^2 .
\ea
\EQN{vac.energy}
\eeq
\end{itemize}
\vspace{3mm}

\noindent
{\it 3. Operator mixing}
\vspace{3mm}

It has been found in \cite{CheSpi88}
that the anomalous dimensions of the
(physical) operators of dimension four  contributing to
\r{142} may be presented
as follows\footnote{
The differences between \r{anom.dim4} and eq.(3.4) of \cite{CheSpi88}
originate from  a different overall sign in the  definition
of $\g_{ij}$ and  the different  normalization of the operator
$Q_1$.}:
\beq
\ba{c}
\dsp
\mu^2\frac{d }{d \mu^2} O_1 =
- \left(a\frac{ \prd}{\prd a}\beta\right) O_1
+ 4\left(a\frac{ \prd}{\prd a}\g_m \right) \sum_i O_2^{ii}
+ 4a\frac{ \prd}{\prd a}\g_0,
\\
\dsp
\mu^2\frac{d }{d \mu^2} O_2^{ij} = - m_{i}
\frac{\prd }{\prd m_j} \g_0,
\\
\dsp
\mu^2\frac{d }{d \mu^2} O_6  = 4\g_m O_6.
\ea
\EQN{anom.dim4}
\eeq
where the last equation is of course a direct consequence
of the definition of the quark mass anomalous dimensions.
\vspace{3mm}

\noindent
{\it 4. RG constraints on the coefficient functions}
\vspace{3mm}

\noindent
In  our particular case of  the polarization operator
$\Pi(Q^2)$ the OPE \r{142} assumes the form
\beq
\ba{c}
\dsp
\Pi(Q^2) \aseqmy{q^2\to\infty}
O(1) + O(m^2/Q^2)
\\
\\
\dsp
+  \frac{\Cbf_1 O_1}{Q^4}
+
\sum_i
 \frac{\Cbf_2^{i,i} O^{ii}_2}{Q^4}
+\frac{\Cbf_2^{ud} O^{ud}_2}{Q^4}
+
\frac{\Cbf_2^{du} O^{du}_2}{Q^4}
+
\frac{\Cbf_6(\as,L,m_i)}{Q^4},
\ea
\EQN{oper}
\eeq
where
\beq
\Cbf_6(\as,L,m_i) =
\sum_{ij}\Cbf_6^{ij} O_6^{ij}
+
\sum_{i}\Cbf_6^{i} O_6^{i}
. 
\EQN{C6}
\eeq
The   condition of the RG invariance  implies
\beq
\dmu \sum_{n} \Cbf_n O_n=0,
\EQN{RG1}
\eeq
or, explicitly,
\beq
\ba{c}
\dsp
 \frac{\prd C_6}{\prd L}
=
- 4\g_m  \Cbf_6
- (a\beta) \frac{\prd \Cbf_6}{\prd a}
\\
\dsp
- \Cbf_1 4a \frac{\prd \g_0}{\prd a}
+  \sum_i \Cbf_2^{ii}  m_i \frac{\prd}{\prd m_i}\g_0
+ \Cbf_2^{ud} m_u \frac{\prd}{\prd m_d} \g_0
+ \Cbf_2^{du} m_d \frac{\prd}{\prd m_u} \g_0 .
\EQN{RG1expl}
\ea
\eeq
The structure of the last equation allows  obviously to find the
logarithmic terms of order $\as^2$ in $\Cbf_6$  in terms of the
function $\Cbf_6$ and  $\Cbf_1$ which are to be known {\em
completely} up to and including the order $\as$ contributions and the
coefficient functions $\Cbf_2^n$ which should be known in the order
$\as^2$. An inspection of \r{144}-\r{146} immediately reveals that
we indeed know these coefficient functions with the accuracy required.
\vspace{3mm}

\noindent
{\it 5. Calculation of VEV's of $O_1$ and $O_2$}
\vspace{3mm}

\noindent
\begin{figure}
\begin{center}
\epsfxsize=4.0cm
\leavevmode
\epsffile{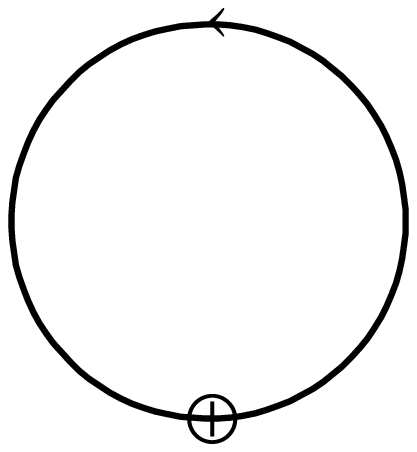}
\epsfxsize=4.0cm
\leavevmode
\epsffile{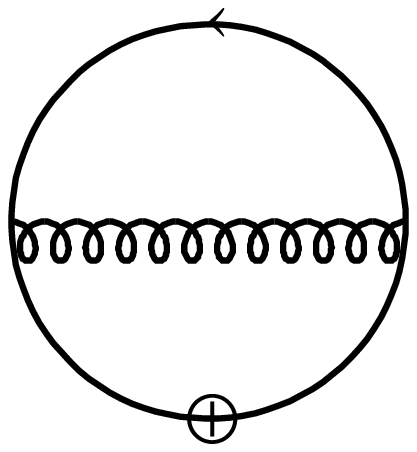}
\epsfxsize=4.0cm
\leavevmode
\epsffile{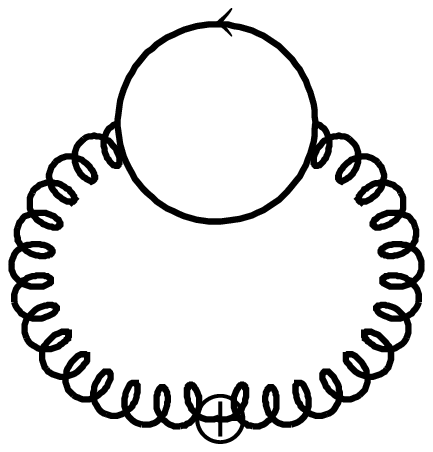}
\caption{\label{diag}Diagrams giving rise to nonzero
vacuum expectation values of the
composite operators $\ovl{\psi} \psi$ and $G^2_{\mu\nu}$ in the lowest
orders of perturbation theory.}
\end{center}
\end{figure}
Diagrams contributing  to VEV's of
$O_1$ and $O_2$  in order $\as$ are shown in Fig.\ref{diag}.
The VEV  of $O_2$  was computed in \cite{Broad,CheSpi88}
and  reads
\beq
<O_2^{ij}>  = \frac{3 m_i m_j^3}{4\pi^2}
\left[
1 + \ln (\frac{\mu^2}{m_j^2}).
+ 2 a(\mu)
\left(
 \ln^2(\frac{\mu^2}{m_j^2})
 + \frac{5}{3}
 \ln (\frac{\mu^2}{m_j^2})
 + \frac{5}{3}
\right)
\right].
\EQN{mqq}
\eeq
The perturbative "gluon condensate" was computed
in \cite{Braat} (eq. (B.1a)):
\beq
<O_1>  =
-\frac{a(\mu)}{2\pi^2}
\sum_i
\left[
9 + 8 \ln (\frac{\mu^2}{m_i^2}) + 3 \ln^2 (\frac{\mu^2}{m_i^2})
\right]
       m_i^4.
\EQN{GG}
\eeq
We have confirmed  the correctness  of the result \r{GG}
by an independent calculation.
\vspace{3mm}

{\it 6. Calculation of the $\as^2 m^4$ corrections}
\vspace{3mm}

\noindent
Now we are in a position to find the corrections
of order $\as^2 m^4/s^2$  to the both spectral functions
$R_V$ and $R_A$. Indeed, simply integrating
the rhs of \r{RG1expl} one gets the missing logarithmic terms of order
$\as^2$ in $\Cbf_6$:
\beq
\ba{l}
\dsp
\sum \fos{T}{C_6} O_6= \\   \dsp
\frac{3}{16 \pi^2}
\left\{
        \left[
         a
      \left(\frac{152}{9} - \frac{32}{3}\zeta(3)\right)
+       a^2
        L\left(
 114
-72 \zeta(3)
+\frac{16}{9} f \zeta(3)
-\frac{76}{27} f
        \right)
        \right]
        m_u^2 m_d^2
\right.
\\
\dsp
       -
        \left[
        2 + a(4 + 4L)
        +a^2
         L \left(
36
+8 L
-\frac{10}{9} f
         \right)
        \right]
      (m_u^4+m_d^4)
  \\
  \dsp
        \mp
        \left[
       4 L+a
       \left(
\frac{56}{3}
-16 \zeta(3)
+\frac{32}{3} L
+8 L^2
       \right)
\right.
\\
\left.
\dsp
-       a^2
        L\left(
\frac{4}{9} f L ^2
+\frac{22}{9} f L
-\frac{8}{3} f \zeta(3)
+\frac{157}{27} f
-18 L ^2
-77 L
+\frac{332}{3} \zeta(3)
-\frac{3617}{18}
         \right)
        \right] \times
\\
\dsp
        (m_u^3 m_d+m_d^3 m_u)
\\
\dsp
  \left.
-       a^2
        L\left(
\frac{2}{3} L
+\frac{16}{3} \zeta(3)
-\frac{40}{9}
	  \right)
         \sum_i m_i^4
\pm       a^2
        16
        L
        m_u m_d  \sum_i m_i^2
\right\}.
\ea
\EQN{C6fin}
\eeq
Note that the term proportional to $\sum_i m_i^4$ in the above
result was independently found via a direct calculation in
\cite{GorSpi85}.

The next step is to use  eqs. (\ref{mqq},\ref{GG},\ref{C6fin})
to find the corrections of order  $\as^2 m_q^4/s^2 $  to
the spectral densities of  the vector and axial vector
correlators. The results reads (below we set for brevity
the $\ovl{\mbox{\rm MS}}$ normalization scale $\mu = \sqrt{s}$
and $\ovl{m}_u(s) = \ovl{m}_d(s) = \ovl{m}(s)$)
\begin{eqnarray}
\dsp
R_V & = & 1 + O(\ovl{m}^2/s)
 -6 \frac{\ovl{m}^4}{s^2}\left(1 + \frac{11}{3} a \right)
\nonumber
\\
&&
\dsp
+
a^2\frac{\ovl{m}^4 }{s^2} \left[
f \left(
\frac{1}{3}  \log\left(\frac{\ovl{m}^2}{s}\right)
-\frac{2}{3}  \pi^2
-\frac{8}{3}  \zeta(3)
+\frac{143}{18}
  \right)
\right.
\nonumber
\\
&&
\left.
\dsp
-\frac{11}{2} \log\left(\frac{\ovl{m}^2}{s}\right)
+27 \pi^2
+112 \zeta(3)
-\frac{3173}{12}
+12  \sum_i \frac{\ovl{m}_i^2}{\ovl{m}^2}
\right.
\nonumber
\\
&&
\left.
\dsp
+
\left(
\frac{13}{3}
-4 \zeta(3)
\right)
\sum_i \frac{\ovl{m}_i^4}{\ovl{m}^4}
-
\sum_i \frac{\ovl{m}_i^4}{\ovl{m}^4}
\log\left(\frac{\ovl{m_i}^2}{s}\right)
\right],
\EQN{VM4}
\end{eqnarray}
\begin{eqnarray}
\dsp
R_A & = & 1 + O(\ovl{m}^2/s)
 +6 \frac{\ovl{m}^4}{s^2}\left(1 + \frac{5}{3} a \right)
\nonumber
\\
&&
\dsp
+
a^2\frac{\ovl{m}^4 }{s^2} \left[
f \left(
-\frac{7}{3}  \log\left(\frac{\ovl{m}^2}{s}\right)
+\frac{2}{3}  \pi^2
+\frac{16}{3}  \zeta(3)
-\frac{41}{6}
  \right)
\right.
\nonumber
\\
&&
\left.
\dsp
+\frac{77}{2} \log\left(\frac{\ovl{m}^2}{s}\right)
-27 \pi^2
-220 \zeta(3)
+\frac{3533}{12}
-12  \sum_i \frac{\ovl{m}_i^2}{\ovl{m}^2}
\right.
\nonumber
\\
&&
\left.
\dsp
+
\left(
\frac{13}{3}
-4 \zeta(3)
\right)
\sum_i \frac{\ovl{m}_i^4}{\ovl{m}^4}
-
\sum_i \frac{\ovl{m}_i^4}{\ovl{m}^4}
\log\left(\frac{\ovl{m_i}^2}{s}\right)
\right].
\EQN{AM4}
\end{eqnarray}
Note that the sum over $i$ includes also the quark
coupled to the external current and with  mass denoted by $m$.
Hence in the case with one heavy quark $u$  of
mass $m$ ($d \equiv u)$ one should set
$\sum_i \frac{\ovl{m}_i^4}{\ovl{m}^4} = 1$
and
$\sum_i \frac{\ovl{m}_i^2}{\ovl{m}^2} = 1$.
In the opposite case when one considers the correlator of light
(massless) quarks the heavy quark appears only
through its coupling to gluons. There one finds:
\beq
\dsp
R_V  =  R_A =
+
a^2 \frac{\ovl{m}^4 }{s^2} \left[
\frac{13}{3}
-\log\left(\frac{\ovl{m}^2}{s}\right)
-4 \zeta(3)
\right].
\EQN{VAM4}
\eeq

Numerically, eqs. (\ref{VM4},\ref{AM4}) look as follows
\begin{eqnarray}
\dsp
R_V & = & 1 + O(\ovl{m}^2/s)
 -6 \frac{\ovl{m}^4}{s^2}\left(1 + \frac{11}{3} a \right)
\nonumber
\\
&&
\dsp
+
a^2\frac{\ovl{m}^4 }{s^2} \left[
f \left(
\frac{1}{3}  \log\left(\frac{\ovl{m}^2}{s}\right)
-1.841
  \right)
\right.
\nonumber
\\
&&
\left.
\dsp
-\frac{11}{2} \log\left(\frac{\ovl{m}^2}{s}\right)
+ 136.693
+12  \sum_i \frac{\ovl{m}_i^2}{\ovl{m}^2}
\right.
\nonumber
\\
&&
\left.
\dsp
 - 0.475   \sum_i \frac{\ovl{m}_i^4}{\ovl{m}^4}
 -
\sum_i \frac{\ovl{m}_i^4}{\ovl{m}^4}
\log\left(\frac{\ovl{m_i}^2}{s}\right)
\right],
\EQN{VM4N}
\end{eqnarray}
\begin{eqnarray}
\dsp
R_A & = & 1 + O(\ovl{m}^2/s)
 +6 \frac{\ovl{m}^4}{s^2}\left(1 + \frac{5}{3} a \right)
\nonumber
\\
&&
\dsp
+
a^2\frac{\ovl{m}^4 }{s^2} \left[
f \left(
-\frac{7}{3}  \log\left(\frac{\ovl{m}^2}{s}\right)
 + 6.157
  \right)
\right.
\nonumber
\\
&&
\left.
\dsp
+\frac{77}{2} \log\left(\frac{\ovl{m}^2}{s}\right)
- 236.515
- 12  \sum_i \frac{\ovl{m}_i^2}{\ovl{m}^2}
\right.
\nonumber
\\
&&
\left.
\dsp
 - 0.475   \sum_i \frac{\ovl{m}_i^4}{\ovl{m}^4}
 -
\sum_i \frac{\ovl{m}_i^4}{\ovl{m}^4}
\log\left(\frac{\ovl{m_i}^2}{s}\right)
\right].
\EQN{AM4N}
\end{eqnarray}

\section{Discussion}
The $Z$ decay rate is hardly affected by the $m^4$ contributions. The
lowest order term evaluated with $\ovl{m}=2.6 \ {\rm GeV}$ leads to a
relative
suppression (enhancement) of about $5\times 10^{-6}$ for the vector
(axial vector) current induced $Z\to b\bar b $ rate.
Terms of increasing order in $\alpha_s$ become successively smaller. It
is worth noting, however, that the corresponding series, evaluated in
the onshell scheme, leads to terms which are larger by about one order
of magnitude and of oscillatory signs. The $m_b^4$ correction
to $\Gamma(Z\to q\bar q)$ which starts in order $\alpha_s^2$
is evidently even smaller.
From these
considerations it is clear that $m^4$ corrections to the $Z$ decay
rate are well under control --- despite the still missing singlet piece
--- and that they can be neglected for all practical purposes.

The situation is different in the low energy region, say
several GeV above the charm or the bottom threshold. For  definiteness
the second case will be considered and for simplicity
all other masses will be put to
zero\footnote{A more detailed analysis of the phenomenological
implications of these results for the low energy region will be
presented in \cite{CK94}.}.
\begin{figure}
\begin{center}
\epsfxsize=13.0cm
\leavevmode
\epsffile[30 130 550 680]{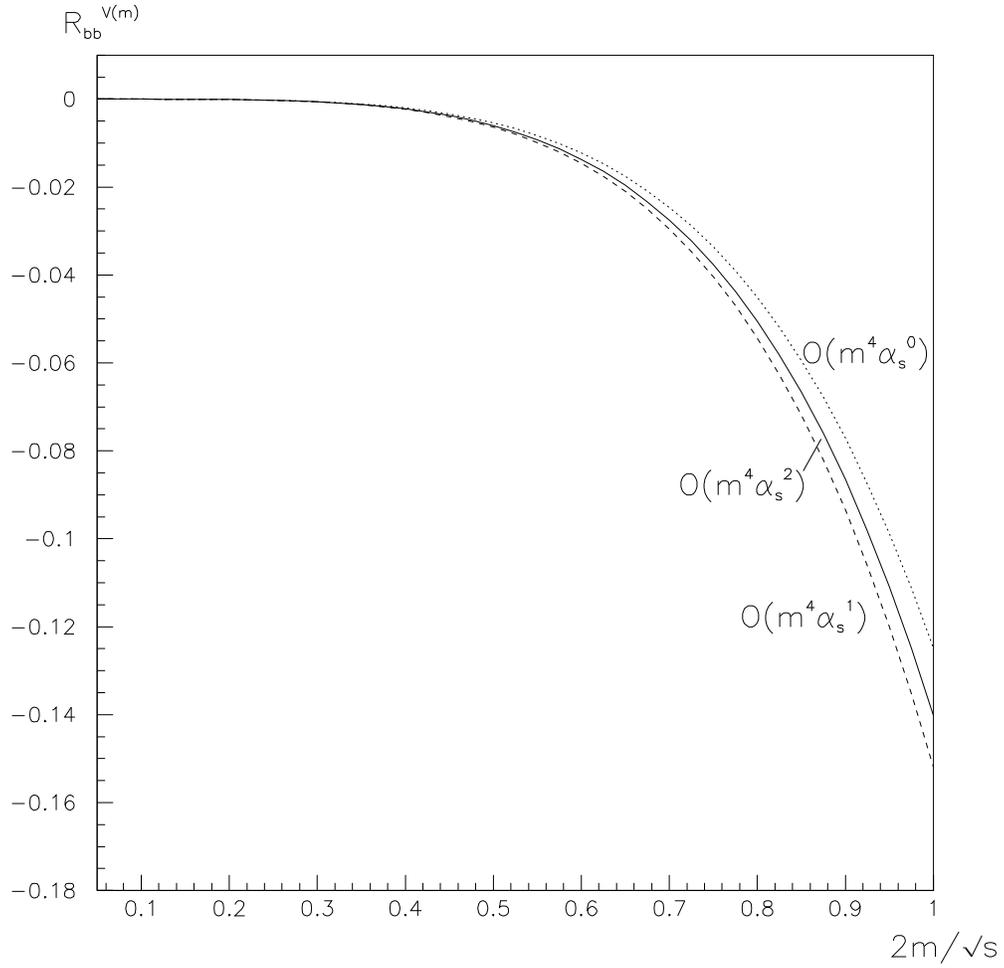}
\caption{\label{alphasexp}Contributions to $R^V$ from $m^4$ terms
including successively higher
orders in $\alpha_s$ (order $\alpha_s^0$/ $\alpha_s^1$/ $\alpha_s^2$
corresponding to dotted/ dashed/ solid lines) as functions of
$2m_{\rm pole}/\protect\sqrt{s}$.}
\end{center}
\end{figure}
The contributions to $R^V$ from $m^4$ terms are
presented  in Fig.\ref{alphasexp}
as functions of $2m/\sqrt{s}$ in the range from 0.05 to 1.
\begin{figure}
\begin{center}
\epsfxsize=13.0cm
\leavevmode
\epsffile[20 150 550 700]{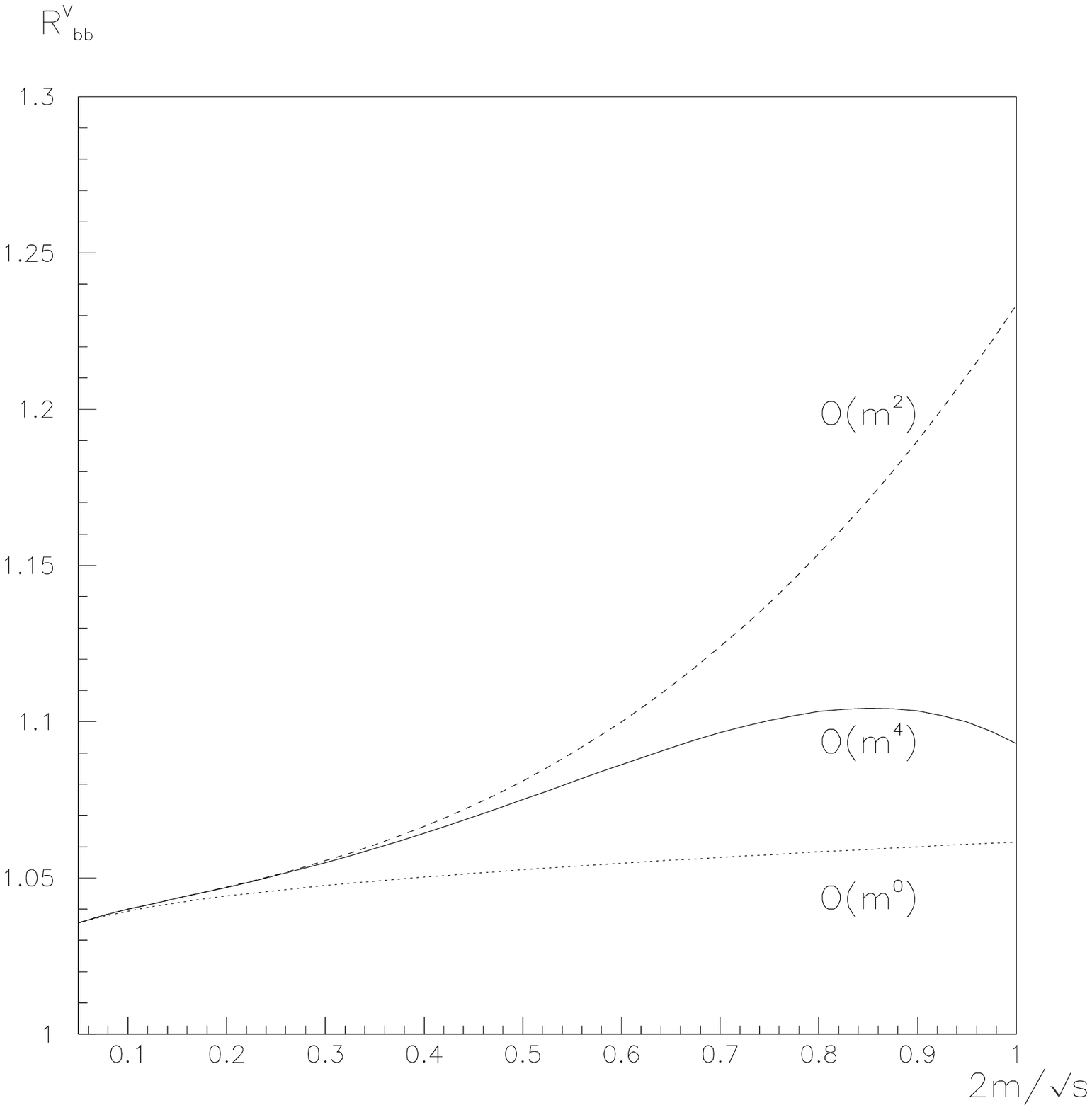}
\caption{\label{massexp}Predictions for $R^V$ including successively higher
orders in $m^2$.}
\end{center}
\end{figure}
As input parameters
$m_{\rm pole}=4.7{\rm GeV}$ and $\Lambda_{\ovl{MS}}=235{\rm MeV}$,
corresponding to $\alpha_s(m_Z^2)=.12$ have been chosen.
Corrections of higher orders are added successively. The prediction is
fairly stable with increasing order in $\alpha_s$ as a consequence of
the fact that most large logarithms were absorbed in the running mass.
The relative magnitude of the sequence of terms from the $m^2$
expansion
is displayed in Fig.\ref{massexp}. The curves for $m^0$ and $m^2$  are based on
corrections up to third order in $\alpha_s$ with the $m^2$ term starting
at first order. The $m^4$ curve receives corrections from order zero to
two.

Of course, very close to threshold, say above 0.75 (corresponding to
$\sqrt{s}$ below 13 GeV) the approximation is expected to break down,
as indicated already in Fig.\ref{kvexp}.
Below the $b \bar b$ threshold, however, one
may decouple the bottom quark and consider mass corrections from the
charmed quark within the same formalism.

Also $R_{q\bar q}$ where $q$ denotes a massless quark  coupled to the
external current is affected by virtual or real heavy quark radiation.
The $m^2$ terms have been calculated in  \cite{ChetKuhn90} and start
from order $\alpha_s^3$:
\beq
\delta R=-\left(\frac{\alpha_s}{\pi}\right)^3
\frac{4\ovl{m}^2}{s} (15-\frac{2}{3} f)(\frac{4}{3} -\zeta(3) )
\eeq
The terms of order $\alpha_s^2m^4$ were given above. Both lead to
corrections of ${\cal O}( 10^{-4})$, (evaluated at an
energy $\sqrt{s}$ of 10 GeV) and can  be neglected for all practical
purposes.
\vspace{3mm}

\noindent{\sl Summary:}
QCD corrections to the vector and axial vector current induced rates
and cross sections of order $\alpha_s m^4$ have  been calculated.
They have little effect on the $Z$ decay rate. However, they are
important for the analysis of hadron production in the low energy
region.
\vspace{5mm}

\noindent{\large \bf Acknowledgement}
We would like to thank A. Kwiatkowski and T. Teubner
for help in preparing the figures.

\end{document}